\documentclass[preprint,11pt]{elsarticle}

\usepackage{amssymb}
\usepackage{graphicx}
\usepackage{physics}
\usepackage{amsmath}
\usepackage{derivative}
\usepackage[version=4]{mhchem}
\usepackage{siunitx}
\usepackage{longtable,tabularx}
\usepackage{floatrow}
\usepackage{url}
\usepackage{hyperref}

\begin{document}

\begin{frontmatter}



\title{Wake Stabilization and Force Modulation via Surface Dimples on an Airfoil at Low-Reynolds-Numbers}


\author[inst1]{Putu Brahmanda Sudarsana}

\affiliation[inst1]{organization={Department of Mechanical Engineering, University of Michigan},
            city={Ann Arbor},
            postcode={48109}, 
            state={MI},
            country={USA}}

\author[inst2]{Jagmohan Singh}
\author[inst1,inst3]{Anchal Sareen}

\affiliation[inst2]{organization={Department of Aerospace Engineering, University of Michigan},
            city={Ann Arbor},
            postcode={48109}, 
            state={MI},
            country={USA}}

\affiliation[inst3]{organization={Department of Naval Architecture and Marine Engineering, University of Michigan},
            city={Ann Arbor},
            postcode={48109}, 
            state={MI},
            country={USA}}

\begin{abstract}
This study investigates the effect of surface dimples on the unsteady aerodynamics of a National Advisory Committee for Aeronautics airfoil (NACA0012) at a chord-based Reynolds numbers of \(Re_c = 5300\) and \(10{,}000\) using direct numerical simulations. Dimples were placed on the suction side at non-dimensional chordwise locations of \(l_D/c\) = 0.035 and 0.35, and the flow response was studied at a fixed angle of attack \(\alpha = 5^\circ\). At \(Re_c = 5300\), dimples placed at \(l_D/c\) = 0.35 reduced lift and drag fluctuations by 26.5\% and 33.3\%, respectively, with minimal change in mean forces. At \(Re_c = 10{,}000\), the same configuration led to a seven-fold increase in force fluctuations, while the mean remained unchanged. The smooth airfoil exhibited irregular, aperiodic force signals at this \(Re_c\), whereas the dimpled case showed highly periodic behavior, indicating wake stabilization. Flow visualizations revealed that dimples generate streamwise vortices within the boundary layer. These vortices are found to have a stabilizing effect on wake dynamics at \(Re_c = 5300\), reducing vortex breakdown and enhancing the coherence of wake structures. Spectral Proper Orthogonal Decomposition (SPOD) showed that dimples redistribute modal energy depending on Reynolds number: at low \(Re_c\), they reduce broadband content and suppress unsteadiness, while at high \(Re_c\), they amplify dominant shedding modes and broaden the spectral energy distribution. These results demonstrate that dimples can passively modulate unsteady forces and wake dynamics for a flow over a streamlined body, either suppressing or enhancing flow instabilities depending on the regime. 
\end{abstract}


\end{frontmatter}

\section{Introduction}\label{sec1}

The past decade has seen rapid advancements in small and micro-unmanned aerial vehicles (UAVs). Their high maneuverability and compact size have expanded their applications to include surveillance, ship decoys, and the detection of biological, chemical, and radioactive materials\cite{Mueller2003}. These UAVs typically feature an airfoil for lift generation and maneuverability. These vehicles operate at low chord-based Reynolds numbers ($Re_c = U_\infty c/\nu \leq 10, 000$, where $c$ is the chord length and $U_\infty$ is the freestream velocity), where the flow is persistently laminar. This limits the airfoil performance compared to the higher Reynolds number \cite{Lissaman1983}. Therefore, a flow control strategy is necessary at low $Re_c$ to enable better performance and maneuverability. One such passive flow control strategy involves surface modifications, such as surface indentations, shaped as dimples, which are widely used in bluff bodies \cite{vilumbrales2025, sudarsana2024, Choi2008}.  Dimples transition the boundary layer near the wall from laminar to turbulent and delay flow separation \cite{Choi2008, vilumbrales2025}. Streamlined bodies, such as airfoils, also exhibit flow separation on the suction side at high angles of attack ($\alpha$). The adverse pressure gradient reduces the flow momentum in the near-wall region and induces instability in the outer boundary layer. It can be reasoned here that dimples will also enable control over the boundary layer transition, which could increase the aerodynamic performance. 

Several experimental and numerical studies have explored the impact of dimples on the aerodynamic performance of airfoils, although at a much larger Reynolds number. Beves et al.\cite{Beves2011} conducted experiments on an inverted Tyrrell026 airfoil with three rows of dimples at $l_D/c=0.23$ ($l_D$ is the distance of the dimple center from the leading edge of the airfoil) to see if they could suppress flow separation at different heights above ground at a Reynolds number ($Re_c$) of $0.5\times10^5$. Their boundary layer profile measurements suggested that the dimples not only induced turbulence in the boundary layer, but also generated strong vortices that helped reattach the flow far downstream, reducing wake thickness and mitigating ground-effect interactions. Miller et al.\cite{Miller2012} conducted a boundary layer analysis on a NACA-2412 airfoil with multiple dimple arrays at the upper leading edge. The study found a positive correlation between dimple depth and the extent of boundary layer attachment. Several studies indicate that the optimal placement for sustaining boundary layer attachment is near the leading edge before the laminar separation point \cite{Xie2022,Ali2024}. However, contrasting findings were reported by Stolt et al.\cite{Stolt2019}, who used Tomographic Particle Image Velocimetry (TPIV) to measure the flowfield around NACA0015 airfoil. They found that a full array of dimples placed at 20\% of the leading edge did not prevent flow separation at $Re_c=1 \times 10^5$. These conflicting results highlight the need for further investigations, and complementary numerical studies could offer a more comprehensive understanding of the effects of dimples on airfoils.

A laminar separation bubble often forms on the suction side of the airfoil, where the laminar boundary layer encounters an adverse pressure gradient, which leads to flow separation, transition to turbulence, and eventual reattachment. This phenomenon can significantly alter the pressure distribution and degrade aerodynamic performance, particularly at low Reynolds numbers. D’Alessandro et al. \cite{DAlessandro2019} conducted a large eddy simulation (LES) of the NACA64(2)-014A airfoil modified with dimples located at $0.55c$ with a dimple depth-to-diameter ratio ($d/D$) of 0.15 at $Re_c=1.75\times10^5$. They reported a significant reduction in the extension of the laminar separation bubble when the dimples were placed very close to the flow separation location. For all three angles of attacks tested in their study ($0^\circ$, $4^\circ$, and $8^\circ$), dimples were observed to reduce pressure drag and increase viscous drag. Subsequent research on the effect of dimples on NLF(1)-0416 airfoil using LES was carried out by Xie et al. \cite{Xie2022}.  They reported drag reductions of up to 43\% and 54\%, along with lift-to-drag ratio improvements of 337\% and more than 280\% in $Re_c=39000$ and $Re_c=68500$, respectively. This was attributed to the formation of a laminar separation bubble due to flow reattachment, in contrast to the smooth baseline case where the flow abruptly separates. A recent study by Ali et al. (2024) used Reynolds-Average Navier-Stokes (RANS) simulations to examine how spherical dimple arrays, both inward and outward, affect the aerodynamic performance of an ONERA M6 wing. They tested the dimples placed at 15\%, 50\%, and 85\% of the wing's leading edge at Reynolds number, $Re_c =  2.5 \times 10^5$. The results showed that the inward dimples on the suction side of the wing reduced the total drag by up to 6.6\% without significantly affecting lift. The researchers attributed this drag reduction mainly to a delay in flow separation caused by the dimples, which primarily affected pressure drag. However, the study did not clarify whether the viscous drag was negatively impacted. Additionally, the use of the RANS method limited the study's ability to predict the complex flow behaviors induced by dimples. 


At ultra-low Reynolds number of $Re_c=5300$, the flow over NACA0012 airfoil is characterized by the absence of any separation bubble and stall condition \cite{Alam2010}. As $\alpha$ increases, the initial fully attached flow becomes partially attached and then abruptly changes to a fully separated shear layer that remains laminar without any signs of reattachment \cite{Alam2010, Wang2014}.
However, for $Re_c \geq 10,000$, the flow is characterized by partially attached laminar boundary layer that reattaches to form a laminar separation bubble. At high angles of attack, the laminar separation bubble bursts leading to a sudden drop of lift, also known as the stall condition \cite{Wang2014}. In this low $Re_c$ regime, small disturbances may cause more than one instability, which could promote transition to turbulence \cite{Gupta2023}. Such disturbances can be in the form of three dimensionality in the wake regime or induced boundary layer perturbation that alters the pressure profile around the airfoil. These other instabilities are typically induced by a perturbed primary instability that eventually alter the steady flow condition into a more complex behavior \cite{Schmid2001}. A study by Gupta et al. \cite{Gupta2023} investigated the onset of three dimensionality on a NACA0012 airfoil using Floquet stability analysis. They found a relation for a critical angle of attack $\alpha_{3D}$ with $Re_c$ as $\alpha_{3D}\sim 1/\sqrt{Re_c}$, with mode C as the first unstable mode for $500\leq Re_c \leq 5000$. The wake regime map in $\alpha-Re_c$ parameter space establish a foundation on the expected wake behavior for the $Re_c\leq5000$. In another study \cite{marquet2022hysteresis} it was found that the flow over a NACA0012 airfoil at $Re_c=5000$ exhibits hysteresis and can display multiple co-existing periodic states when the angle of attack is varied between $7^\circ$ and $8^\circ$. Specifically, the flow can settle into different saturated unsteady states-characterized by distinct wake patterns and lift signals-depending on whether the angle of attack is being increased or decreased, including the occurrence of period-doubling bifurcations. 

Small and micro UAVs routinely operate at ultra-low to low chord-based Reynolds numbers ($Re_c \lesssim 10^4$), where flow separation, transition, and unsteady aerodynamic forces are particularly sensitive to surface perturbations. Despite the widespread application of surface dimples in higher-Reynolds-number flow control, their influence in this low-$Re$ regime remains largely unexplored. This study aims to address that gap. Using direct numerical simulations (DNS), we investigate the effects of surface dimples on the aerodynamic performance and flow structures of a NACA0012 airfoil at $Re_c = 5{,}300$ and $10{,}000$. We hypothesize that dimples placed near the natural separation point can generate streamwise vortices, promote boundary layer transition, and potentially induce reattachment by energizing the near-wall flow. However, as this study will reveal, the effects of dimples are far more multifaceted, striking, and strongly Reynolds-number dependent than initially anticipated---making this an especially compelling case for further exploration.

\section{Methodology}\label{sec2}

\subsection{\label{sec2sub1}Numerical Methods}
The incompressible Navier--Stokes equations for a Newtonian fluid in their non-dimensional form are written as:
\begin{align}
\label{ns1}
    \pdv{u_i}{x_i} & = 0 \\ \nonumber
    \pdv{u_i}{t}+u_j \pdv{u_i}{x_j} & = -\pdv{p}{x_i} + \frac{1}{Re_c}\pdv {u_i}{x_j,x_j}, 
\end{align}

where $u_i$ is the velocity in $i$-direction, $p$ is the non-dimensional pressure and $Re_c$ is the chord-based Reynolds number, defined as $Re_c=U_\infty c/\nu$ where $U_\infty$ is the free-stream velocity, $c$ is the chord length, and $\nu$ is kinematic viscosity of the fluid. Eq.~\ref{ns1} is non-dimensionalized using the length scale $c$, velocity scale $U_\infty$ and time scale $c/U_\infty$. These continuity and momentum equations are solved using an \texttt{OpenFOAM v2406} solver, \emph{pimpleFoam}, which is a transient solver for the incompressible flow. It solves Navier--Stokes equations using a combination of the Pressure Implicit with Splitting of Operator (PISO) algorithm, proposed by Issa \cite{Issa1986}, and the Semi-Implicit Method for Pressure Linked Equations (SIMPLE) by Patankar and Spalding \cite{Patankar1972}, which enables the pressure-velocity coupling in the predictor and corrector steps that satisfy continuity and provide better stability at higher time steps (for more details of the numerical method, see \cite{openfoam2406}). The time derivative term in Eq.~\eqref{ns1} is discretized using the second-order accurate implicit Backward Euler method. The gradient, divergence, and Laplacian operators are discretized using the finite volume Gaussian integration with a linear interpolation scheme, providing second-order accuracy in space. The surface normal gradient operator is calculated using the limited scheme which combines the corrected and uncorrected schemes with a factor of 0.5. In the predictor step, the equation is solved using the Preconditioned Bi-Conjugate Gradient (PBiCG) method, preconditioned with a Diagonal Incomplete-Lower-Upper (DILU) decomposition. In the corrector step, the equation is solved using the Preconditioned Conjugate Gradient (PCG) method, preconditioned with the Diagonal Incomplete Cholesky (DIC) decomposition. Both iterative procedures are set with a $1 \times 10^{-6}$ tolerance. The current solver configuration yields to time step continuity error on the order of $1 \times 10^{-12}$ by constraining the CFL (Courant–Friedrichs–Lewy) number at 0.8.

\subsection{\label{sec2sub2}Simulation Setup and Computational Details}
\begin{figure}[htbp]
  \centerline{\includegraphics[scale = 0.14]{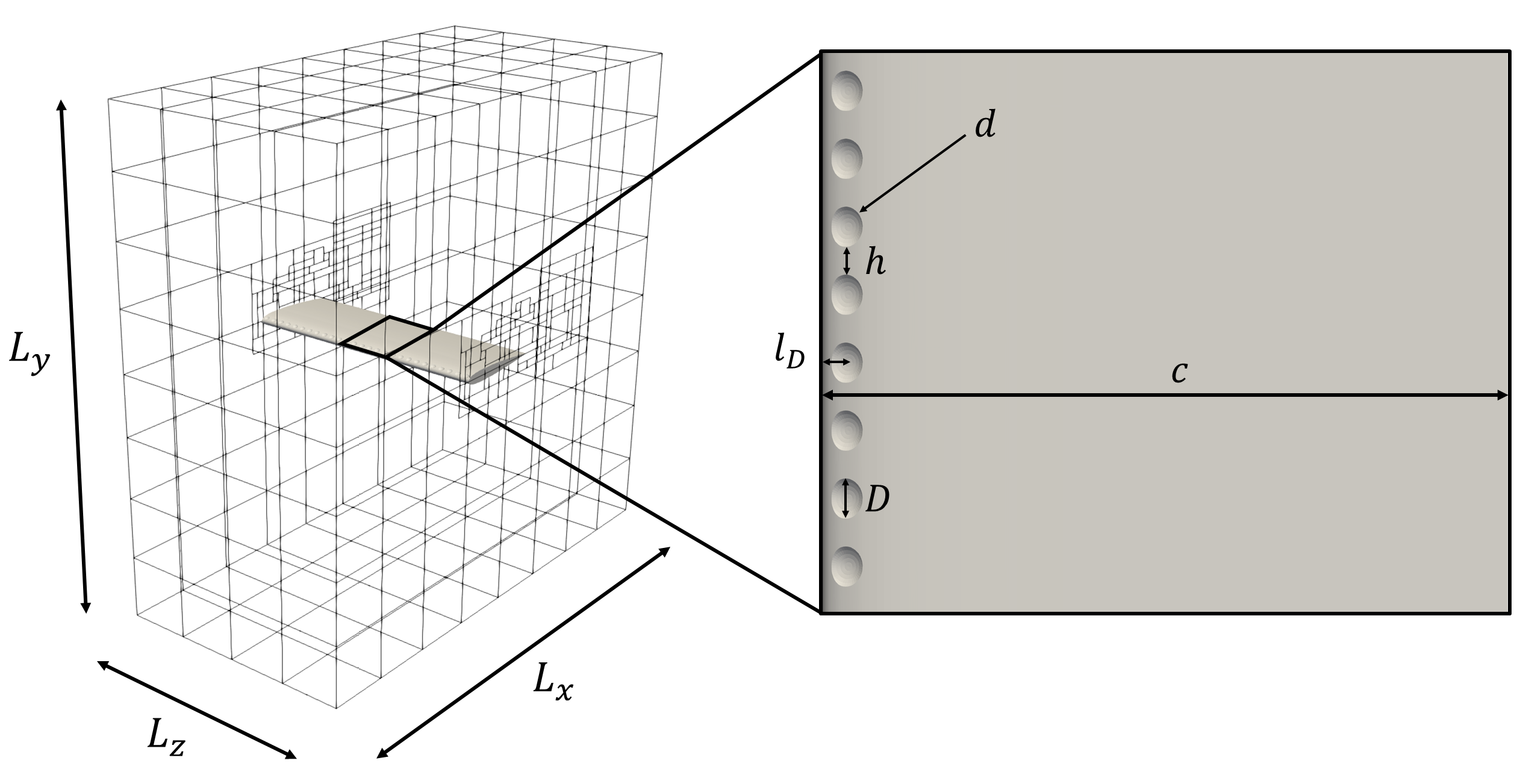}}
  \caption{Computational domain with dimpled NACA-0012 airfoil. Geometric details of the dimple array is shown on the right. Here, $L_x$, $L_y$ and $L_z$ are the dimensions of the computational box, $d$ is the dimple depth, $D$ is the dimple diameter, $l_D$ is the distance of the center of the dimple array to the airfoil leading edge, $h$ is the spacing between the dimples. $c$ and $s$ refer the chord and span of the airfoil, respectively.}
\label{fig1}
\end{figure}

Figure~\ref{fig1} shows the computational domain, which is a box with dimensions ($L_x, L_y, L_z$) = ($10.8c, 5.4c, 2.7c$), where $x$, $y$, and $z$ represent the streamwise, wall-normal, and spanwise directions, respectively. Flow past a symmetric NACA0012 airfoil with an aspect ratio, defined as the ratio of span to chord length ($AR=s/c$) of 2.7 is considered. Airfoil walls are modeled as no-slip walls. The angle of attack $\alpha$ is specified via the inlet flow angle. Thus, the lower $x$ and lower $y$ faces become the inlet for the computational domain where Dirichlet and Neumann boundary conditions (BCs) are imposed for velocity and pressure, respectively. Similarly, the outlet is in the opposite normal direction of the inlet section, where Dirichlet and Neumann BCs are used for pressure and velocity, respectively. 

The airfoil aspect ratio is selected based on previous experimental studies on the same foil profile \cite{Wang2014, Domel2018a}. For the simulations with dimples, the airfoil is featured with an array of dimples on the suction side as shown on the right side of Fig.~\ref{fig1}. Other than chord length $c$, geometric parameters for this configuration are the dimple spacing, $h$, the dimple depth, $d$, dimple diameter, $D$, and the location of the dimple from the leading edge, $l_D$. The dimple spacing is set to be half of the dimple diameter, $h=D/2$ whereas the dimple depth-to-diameter ratios of $d/D$ = 0 (smooth configuration) and 0.2 are considered. Previous studies follow a design criteria of dimple with $d/D=0.1$ up to $d/D=0.15$ located in front of the separation bubble with depth-to-boundary layer thickness ratio 0.6 to 1 for achieving flow separation control and minimal pressure losses \cite{robarge2004design,Xie2022}. The current study refers to the criteria of the ratio of perturbation depth-to-boundary layer thickness of $\approx1$ with $d/D=0.2$. The boundary layer thickness that is used for reference is at $x/c=0.035$ up to $x/c=0.2$.

Two sets of dimple array locations from the airfoil leading edge ($l_D$) are considered, one close to the leading edge, and the other based on the flow separation location identified in the smooth airfoil simulations. Both locations are considered due to the instabilities exhibited in non-dimpled airfoil generally located at these points, i.e. the shear layer instability and the laminar separation bubble at $Re_c=5300$ and $Re_c=10,000$ \cite{Alam2010, Wang2014}. Three-dimensional disturbances induced by dimples are hypothesized to affect the overall flow structure and behavior in the present study. A moderate angle of attack of $\alpha=5^\circ$ is considered. A detailed list of test parameters used in the simulation is shown in Table~\ref{tabpar}.

\begin{table}[htbp]
  \caption{List of test parameters for the simulations. $Re_c$ is the chord-based Reynolds number, $\alpha$ is the airfoil angle of attack, $d/D$ is the dimple depth to diameter ratio, and $l_D$ is the dimple array location}
  \begin{tabular}{lccc}
      \hline 
      $Re_c$          & $\alpha$ $(^o)$ &   $d/D$     & $l_D/c$ \\
      \hline
       $1000$~~~~~~~    & ~~~10~~  & ~~~0~~~~~  & ~~-~~ \\
       $5300$  & ~~~5~~~  & ~~~0~~~~ & ~~-~~ \\
                        & ~~~5~~~  & ~~~0.2~~~~ & ~~$0.035$ \\
                        & ~~~5~~~  & ~~~0.2~~~~ & ~~$0.35$\\
       $10000$~~~~~~~    & ~~~5~~~  & ~~~0~~~~ & ~~-~~ \\
                        & ~~~5~~~  & ~~~0.2~~~~ & ~~$0.035$ \\
                        & ~~~5~~~  & ~~~0.2~~~~ & ~~$0.35$\\
        \hline
  \end{tabular}
  \label{tabpar}
\end{table} 

The computational domain is discretized using finite volume hexahedral cells generated using the \texttt{OpenFOAM} meshing utilities, the \texttt{blockMesh} and \texttt{snappyHexMesh}. In total, there are approximately 9.8 million cells in the medium mesh and 26.7 million cells in the fine mesh with the near-wall resolution $\Delta y^+_{min} = 0.033$ in both. Here $\Delta y^+$ is the cell height near the airfoil surface defined in viscous units as $\Delta y^+ = \Delta y u_\tau/\nu$ with $u_\tau =\sqrt{\tau_w/\rho}$ and $\tau_w=\mu \frac{\partial u}{\partial y}\Bigr|_{\substack{y=0}}$. Both configurations consist of two box refinement regimes, one around the airfoil and another in the rear wake regime, from $x=1.8c$ to $x=2.7c$. The main difference between the two mesh configurations lies in the rear wake regime, where the medium mesh configuration is coarser compared to the fine configuration. Mesh convergence is verified using the lift and drag coefficients at $Re_c=5300$ as shown in Table \ref{tabpar0}. The force coefficients changed only marginally between the medium and fine mesh confirming the mesh convergence. All subsequent analyses used the fine mesh configuration. 

\begin{table}[htbp]
  \caption{Mesh convergence analysis at $Re_c=5300$}
  \begin{tabular}{llcc}
  \hline
      Category & Mesh Count   & $C_L$ & $C_D$ \\
      \hline
       Medium & 9,893,620~   & 0.2135 & 0.0668 \\
       Fine & 26,772,923   & 0.2177 & 0.0671 \\
       \hline
  \end{tabular}
  \label{tabpar0}
\end{table} 

\subsection{\label{sec2sub3}Verification and Validation}
\begin{figure}[htbp]
  \centerline{\includegraphics[scale = 0.35]{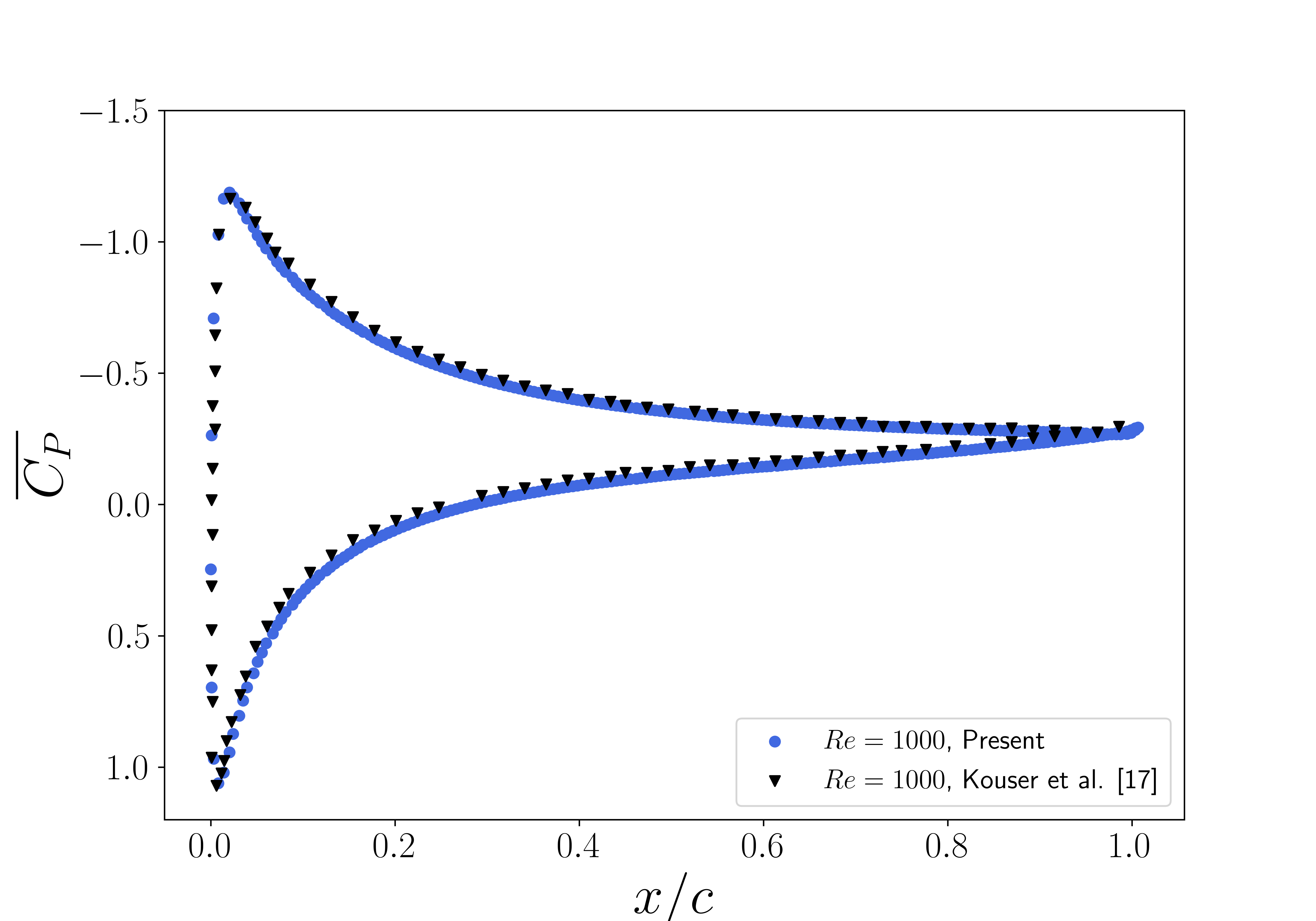}}
  \caption{Comparison of the time-averaged pressure coefficient at a midspan plane, $\overline{C_P}$ with previous simulation study at $Re=1000$ and $\alpha=10^{\circ}$}
\label{fig2}
\end{figure}

The numerical approach is verified using a smooth airfoil configuration by comparing the present results with those available in the literature, as summarized in Table~\ref{tabpar1} and Table~\ref{tabpar2}. Table \ref{tabpar1} presents the comparison of mean lift coefficient $\overline{C_L}=2 \overline{F_L}/\rho U_{\infty}^2 c s$, mean drag coefficient $\overline{C_D}=2 \overline{F_D}/\rho U_{\infty}^2 c s$, and the Strouhal number, $St=fc/U_\infty$ at $Re_c=1000$ with results from previous studies employing different numerical methods. The 2D simulation on NACA0012 airfoil by Mittal and Tezduyar \cite{Mittal1994} was conducted using finite element formulation, while Falagkaris et al. \cite{Falagkaris2017} and Fang et al. \cite{Fang2019} used lattice Boltzmann method. In contrast, Kouser et al. \cite{Kouser2021} conducted a 3D simulation of NACA0012 airfoil using the finite volume method, which is similar to the approach used in the current study with second-order discretization in space and time. The current simulations employ slip-wall boundary conditions at the side walls to represent a finite wing, whereas the 3D simulations by Kouser et al. \cite{Kouser2021} imposed periodic boundary conditions, effectively modeling an infinite wing. As shown in Table~\ref{tabpar1} and Table~\ref{tabpar2}, the results from the current study are in good agreement with previous work. Figure~\ref{fig2} presents the mean pressure coefficient profile, $\overline{C_p} = 2\overline{p}/\rho U_{\infty}^2$, at the foil midspan, which closely matches the data reported in \cite{Kouser2021}. Additionally, the solution was verified to be grid-converged, as further mesh refinement did not alter the results. These findings confirm that the current numerical setup accurately captures the flow physics and reproduces key aerodynamic quantities, demonstrating strong validation of the simulation framework.


\begin{table*}
\caption{\label{tabpar1}Validation of the smooth NACA0012 airfoil result with previous simulation studies at $Re_c=1000$ and $\alpha=10^{\circ}$}
\begin{tabular}{lccccc}
\hline
      ~~~~~~~~ & Present &   Kouser,  & Fang,   & Falagkaris, & Mittal and \\
      ~~~~~~~~ &    &  et al. \cite{Kouser2021}  &  et al. \cite{Fang2019}  &  et al. \cite{Falagkaris2017}  &  Tezduyar \cite{Mittal1994} \\
      \hline
       Mean lift coefficient $\overline{C_L}$  & 0.4367 & 0.4152 & 0.4203 & 0.4178 & 0.4221 \\
       Mean drag coefficient $\overline{C_D}$  & 0.1694 &  0.1656 & 0.1674 & 0.1667 & 0.1650 \\
       Strouhal Number $St$  & 0.875 & 0.869 & - & 0.861 & 0.862 \\
       \hline
\end{tabular}
\end{table*}

Further validation of the current numerical approach is conducted at a higher Reynolds number of $Re_c = 5300$ by comparing the results with experimental data from \cite{Wang2014} (Table \ref{tabpar2}). The experiments were performed for a NACA0012 airfoil with an aspect ratio of $s/c = 2.7$ in a water channel with a turbulence intensity of $T_u = 0.6\%$. The current results show good agreement with the experimental data, with small deviations in $\overline{C_L}$ (2.4\%) and $\overline{C_D}$ (4.2\%). These discrepancies may be attributed to uncertainties in the experimental measurements, which is 5\% as reported in ref. \cite{Wang2014}, and the absence of freestream turbulence in the simulations.


\begin{table}[htbp]
  \caption{Validation of the smooth NACA0012 airfoil result with previous experimental study at $Re_c=5300$ and $\alpha=5^{\circ}$}
  \begin{tabular}{llllll}
  \hline
      ~~~~~~~~ & Present &   Wang et al. \cite{Wang2014}  \\
      \hline
       Mean lift coefficient $\overline{C_L}$  & 0.2177 & 0.2125 \\
       Mean drag coefficient $\overline{C_D}$  & 0.0671 & 0.0701 \\
       \hline
  \end{tabular}
  \label{tabpar2}
\end{table}

\section{Results and Discussions}\label{sec3}

\subsection{\label{sec3sub1}Aerodynamics performance}
The effect of dimples is analyzed on the mean lift and drag coefficient $\overline{C_L}$ and $\overline{C_D}$, and the root mean square (RMS) of the lift and drag fluctuations $C'_{L,rms}$ and $C'_{D,rms}$ at $Re_c=5300$ and $Re_c=10,000$. The $\overline{C_L}$ and $\overline{C_D}$ for $Re_c=5300$ and $Re_c=10,000$ are time-averaged over $t^*=10$ and $t^*=20$ respectively, where $t^*=tU/c$ denotes the flow through time based on the airfoil chord. For both Reynolds numbers, the presence of dimples at $l_D/c=0.035$ and $l_D/c=0.35$ does not result in a significant change in $\overline{C_L}$ and $\overline{C_D}$, as the variations remain within 3\% compared to the smooth airfoil (Tables \ref{tabpar3} and \ref{tabpar4}). This behavior differs from previous studies conducted at higher Reynolds numbers ($Re_c > 50,000$) \cite{Xie2022, Ali2024}, where a notable improvement in aerodynamic performance was reported, particularly in drag reduction due to delayed global flow separation. At higher Reynolds numbers, the boundary layer flow is more susceptible to finite-amplitude perturbations, promoting the transition to turbulence. Additionally, in that regime, flow instability features such as laminar separation bubbles have been observed \cite{eljack2021, Xie2022}. In contrast, for the current study at $Re_c \leq 10,000$ and an angle of attack of $\alpha = 5^\circ$, the flow remains persistently laminar. It is therefore suspected that the dimples at both locations are not effective in triggering boundary layer transition. 

Although no significant changes are observed in $\overline{C_L}$ and $\overline{C_D}$, a reduction in force fluctuations is observed for the dimpled airfoil at $l_D/c=0.35$ at $Re_c=5300$. Compared to the smooth airfoil, the dimpled airfoil $l_D/c=0.35$ reduces lift coefficient fluctuations $C'_{L,rms}$ by 26.5\% and drag coefficient fluctuations $C'_{D,rms}$ by 33.3\% (Table \ref{tabpar3}). Conversely, no noticeable effect on force fluctuations is observed for the dimpled airfoil at $l_D/c= 0.035$. Experimental results by Wang et al. \cite{Wang2014} suggest that at $Re_c=5300$ and $\alpha=5^\circ$, the leading-edge flow remains laminar, with partial separation occurring along the airfoil body. This supports the present findings, as dimples at $l_D/c=0.035$ do not induce sufficient flow perturbations to alter aerodynamic behavior.

\begin{figure*}[htbp]
  \centerline{\includegraphics[scale = 0.42]{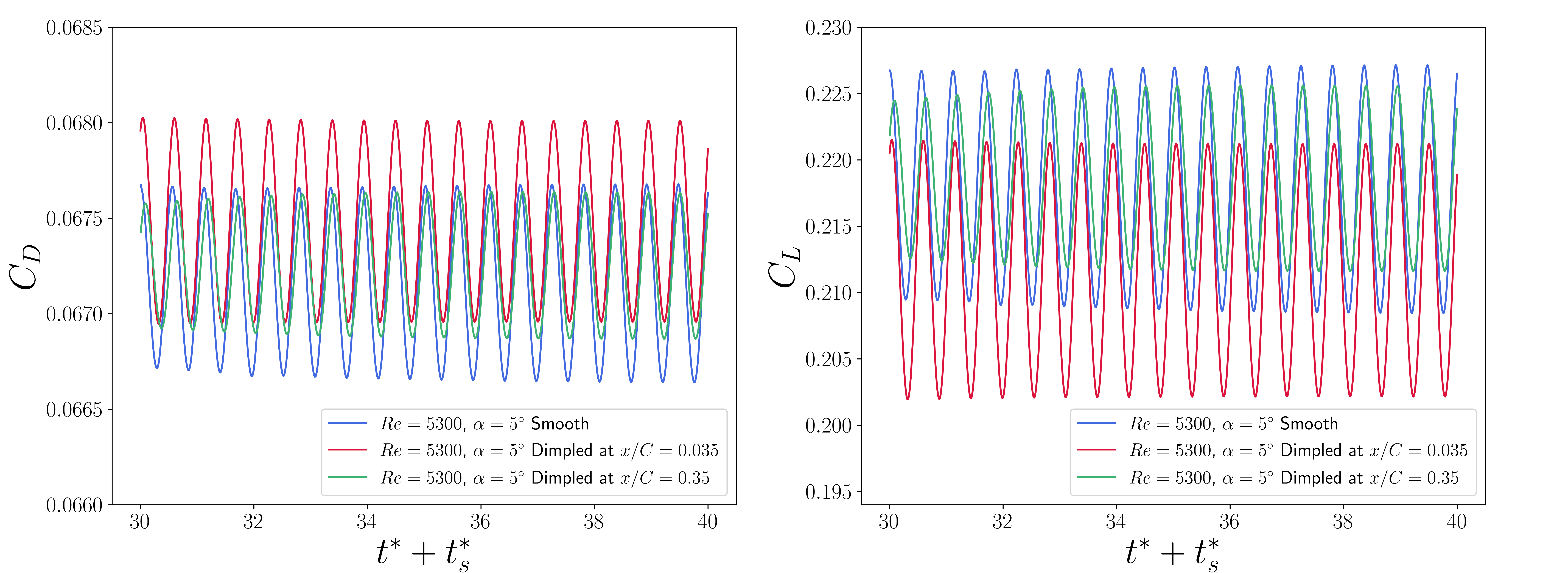}}
  \caption{Temporal evolution of $C_D$ and $C_L$ at $Re=5300$ and $\alpha=5^{\circ}$. Here,  $t^*=tU/c$ denotes the flow through time based on the airfoil chord and $t_s^*$ is a time-shifted parameter introduced to facilitate comparison. }
\label{fig3} 
\end{figure*}

The reduction in force fluctuations for the dimpled airfoil $l_D/c=0.35$ is further shown in the temporal evolution of $C_L$ and $C_D$ (Fig. \ref{fig3}). To facilitate direct comparison, a time-shifted parameter $t^*_s$ is introduced along the $x$-axis, assuming that the fluctuating behavior has reached a statistically steady state. The results show a noticeable reduction in the amplitude of the fluctuation for the dimpled airfoil at $l_D/c=0.35$ compared to both the smooth and the $l_D/c=0.035$ dimpled cases. Similar suppression of force fluctuations has been reported in bluff-body flows through boundary layer perturbations \cite{vilumbrales2025, sudarsana2024, Sareen2024}, although at a much higher Reynolds number ($Re\geq 90,000$). These findings suggest that dimples stabilize the shear layer and suppress unsteady aerodynamic forces, without significantly altering the mean separation location.

\begin{table*}[htbp]
  \caption{Comparison between smooth and dimpled ($d/D=0.2$) NACA0012 airfoil at $Re_c=5300$ and $\alpha=5^{\circ}$}
  \begin{tabular}{lllll}
  \hline
      ~~~~~~~~ & Smooth & Dimpled & Dimpled \\
      ~~~~~~~~ &  & at $l_D/c=0.035$ & at $l_D/c=0.35$ \\
       \hline
       Mean lift coefficient $\overline{C_L}$  & 0.2177 & 0.2115 & 0.2185  \\
       Mean drag coefficient $\overline{C_D}$  & 0.0671 & 0.0674 & 0.0672  \\
       RMS of lift coefficient fluctuations $C'_{L,rms}$  & 0.0064 & 0.0067 & 0.0047 \\
       RMS of drag coefficient fluctuations $C'_{D,rms}$  & 0.0003 & 0.0003 & 0.0002 \\
       \hline
  \end{tabular}
  \label{tabpar3}
\end{table*}

\begin{figure*}[htbp]
  \centerline{\includegraphics[scale = 0.42]{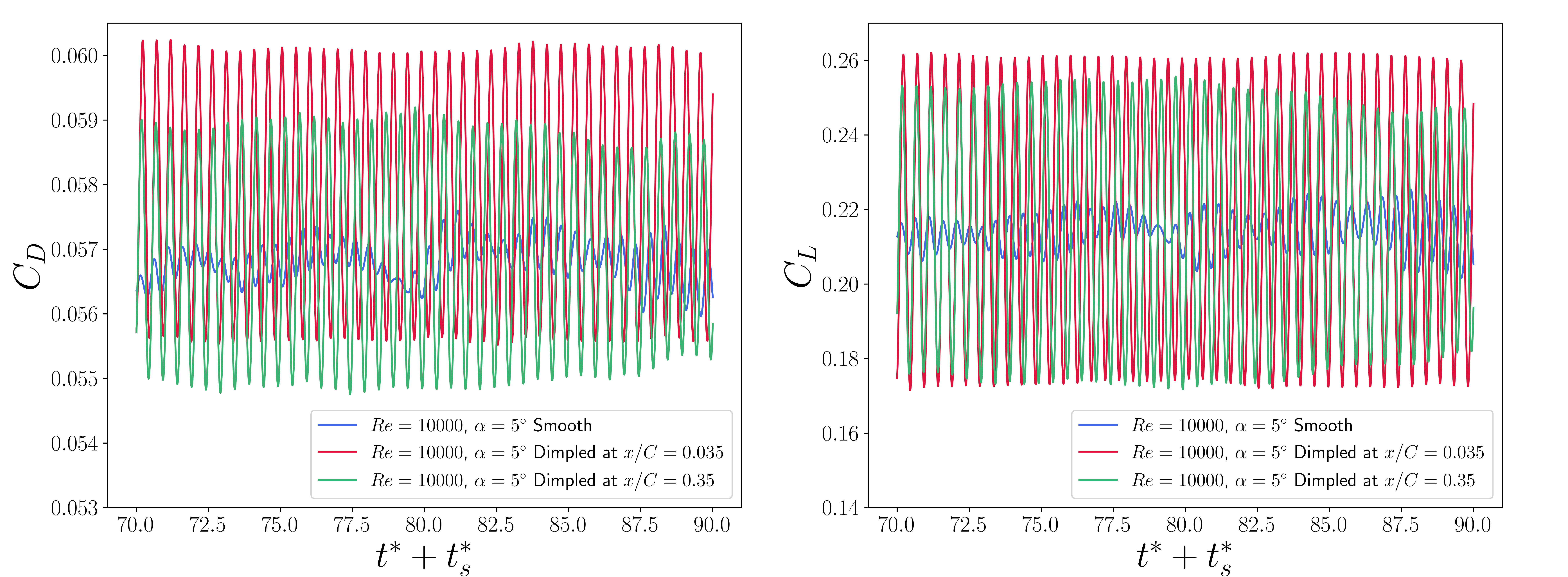}}
  \caption{Temporal evolutions of $C_D$ and $C_L$ at $Re=10,000$ and $\alpha=5^{\circ}$. }
\label{fig4}
\end{figure*}

The opposite behavior is observed at $Re_c=10,000$ where both cases of the dimpled airfoil at $l_D/c=0.035$ and $l_D/c=0.35$ exhibit larger force coefficient fluctuations compared to the smooth case. The dimple airfoil $l_D/c=0.035$ shows a 576\% increase in $C'_{L,rms}$ and a sevenfold increase in $C'_{D,rms}$ relative to the smooth airfoil. Similarly, the dimpled airfoil $l_D/c=0.35$ experiences slightly smaller fluctuations than the $l_D/c=0.035$, but these fluctuations remain significantly higher than the smooth airfoil, with a 489\% larger $C'_{L,rms}$ and a sixfold large $C'_{D,rms}$ compared to the smooth airfoil. Interestingly, there are no significant differences in the mean lift or drag coefficients. Although the overall fluctuations of the smooth airfoil remain lower than those of the dimpled cases, an intriguing observation is seen in the temporal evolution of the force coefficients at $Re_c=10,000$. The smooth airfoil exhibits an irregular sinusoidal pattern (Fig. \ref{fig4}), while the two dimpled airfoils display a highly periodic sinusoidal pattern with predictable maximum and minimum fluctuations. This erratic behavior in the force coefficients of the smooth airfoil may be related to the transition to turbulence within the boundary layer and the wake region. Previous experimental studies have indicated that flow over a NACA0012 airfoil at $Re_c=10,000$ exhibits pronounced transitional behavior and suggested $Re_c=10,000$ as the boundary between the ultra-low and low Reynolds number regimes \cite{Alam2010, Wang2014}. Under these conditions, the presence of a dimple array on the airfoil may influence the transition process, potentially delaying its onset at this Reynolds number. Further analysis of the underlying mechanisms is presented in section~\ref{sec3sub2}.

\begin{table*}[htbp]
  \caption{Comparison between smooth and dimpled ($d/D=0.2$) NACA0012 airfoil at $Re_c=10,000$ and $\alpha=5^{\circ}$}
  \begin{tabular}{llll}
  \hline
      ~~~~~~~~ & Smooth & Dimpled  & Dimpled  \\
      ~~~~~~~~ &  & at $l_D/c=0.035$ & at $l_D/c=0.35$ \\
       \hline
       Mean lift coefficient $\overline{C_L}$  & 0.2144 & 0.2162 & 0.2133 \\
       Mean drag coefficient $\overline{C_D}$  & 0.0568 & 0.0577 & 0.0568 \\
       RMS of lift coefficient fluctuations $C'_{L,rms}$  & 0.0046 & 0.0311 & 0.0271 \\
       RMS of drag coefficient fluctuations $C'_{D,rms}$  & 0.0002 & 0.0016 & 0.0014 \\
       \hline
  \end{tabular}
  \label{tabpar4}
\end{table*}

\begin{figure*}[htbp]
  \centerline{\includegraphics[scale = 0.42]{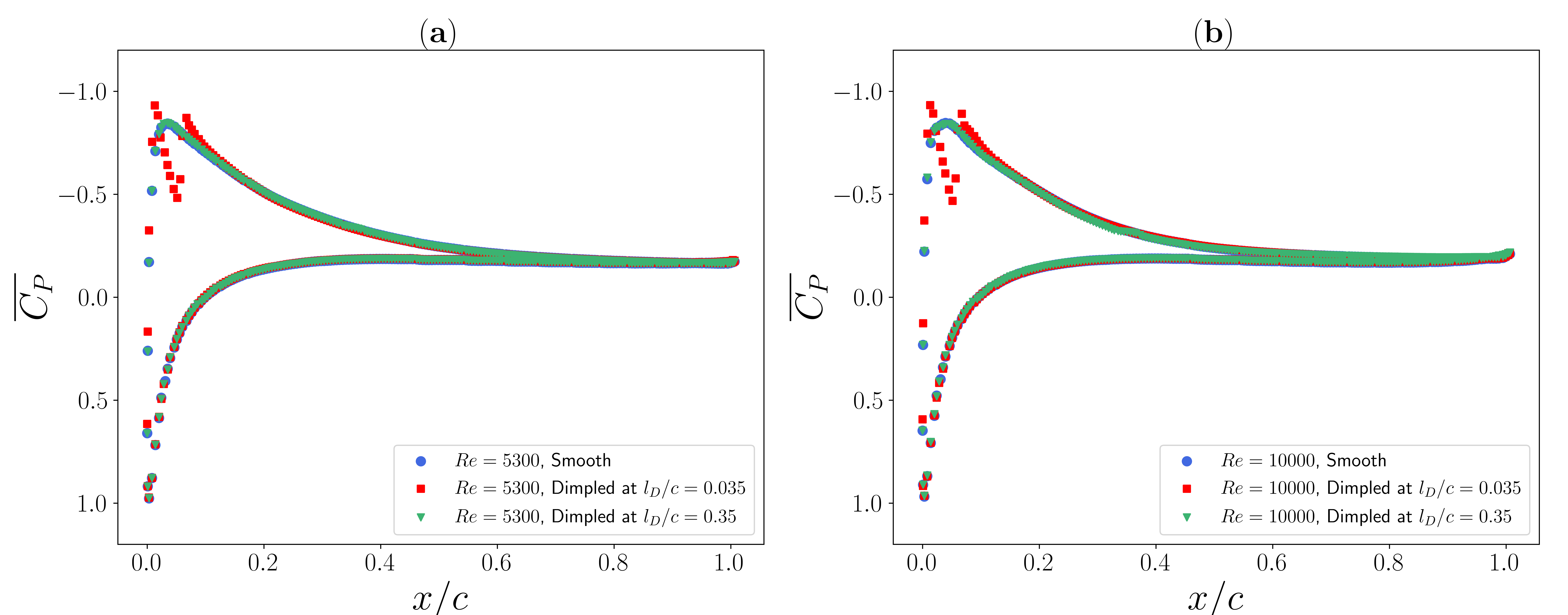}}
  \caption{Time-averaged pressure coefficient, $\overline{C_P}$ at (a) $Re=5300$, $\alpha=5^{\circ}$ and (b) $Re=10,000$, $\alpha=5^{\circ}$}
\label{fig5}
\end{figure*}

Figure \ref{fig5} shows the time-averaged pressure coefficient profile $\overline{C_P}$ at the mid-span, $L_z/2$, for all cases. Consistent with the trends observed in the mean lift and drag coefficients, $\overline{C_L}$ and $\overline{C_D}$, no significant change in the overall pressure coefficient profile is observed by addition of dimples for both $Re_c=5300$ and $Re_c=10,000$. The dimple array placed at $l_D/c=0.35$ produces a profile nearly identical to that of the smooth airfoil, while minor variations are observed when the dimples are placed near the leading edge at $l_D/c=0.035$. However, these deviations diminish rapidly as $x/c$ increases. These changes could be attributed to the presence of alternating adverse pressure gradient ($dp/dx>0$) and favorable pressure gradient ($dp/dx<0$) that develop over a small distance in $x$-direction due to the dimple array at the leading edge. This is observed for both $Re_c=5300$ and $Re_c=10,000$, however, since these deviations in the pressure coefficient profile develop only over a short distance, their overall effect on aerodynamic performance remains negligible, as reflected in the consistent values of $\overline{C_L}$ and $\overline{C_D}$ for different cases. 

\subsection{\label{sec3sub2}Flow structures and Vortex Shedding}
\begin{figure*}[htbp]
  \centerline{\includegraphics[scale = 0.26]{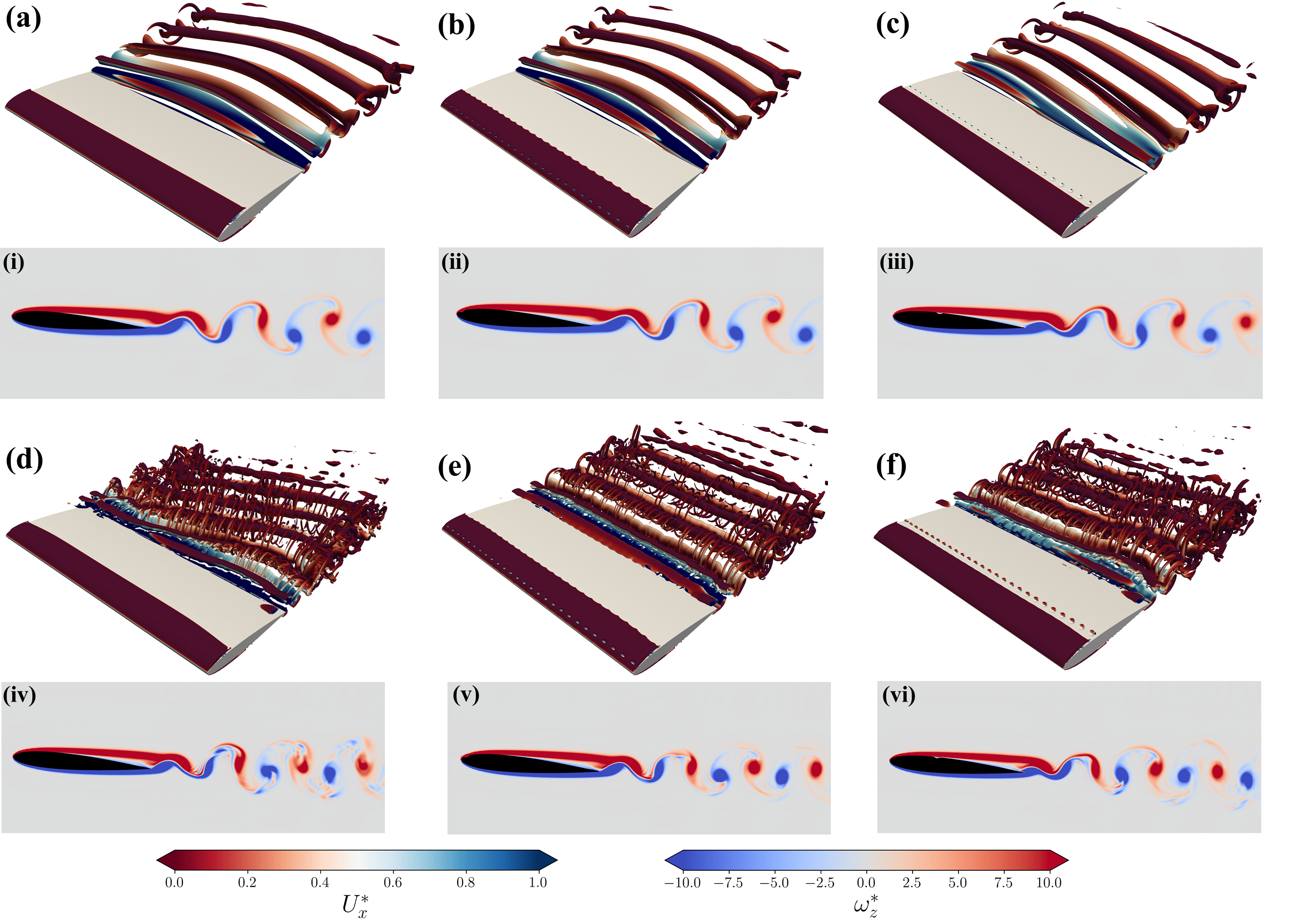}}
  \caption{Instantaneous iso-surfaces at $Q=8$  (a)$-$(f) and the corresponding side view at $L_z/2$ of instantaneous normalized spanwise vorticity $\omega_z^*$ (i)$-$(vi) at $Re_c=5300$ ((a)$-$(c), (i)$-$(iii)) and  $Re_c=10,000$ ((d)$-$(f), (iv)$-$(vi)) where (a) \& (i) smooth airfoil at $t^*=30$, (b) \&(ii) dimpled airfoil at $l_D/c=0.035$ at $t^*=40$, (c) \& (iii) dimpled airfoil at $l_D/c=0.35$ at $t^*=40$, (d)\& (iv) smooth airfoil at $t^*=90$, (e) \& (v) dimpled airfoil $l_D/c=0.035$ at $t^*=50$, and (f) \& (vi) dimpled airfoil $l_D/c=0.35$ at $t^*=90$. The $Q$ iso-surface is overlaid with the color bar of the instantaneous normalized streamwise velocity $U_x^*$ (left), while the color bar of the spanwise vorticity $\omega_z^*$ is shown on the right side. The black color shows the airfoil body.}
\label{fig6}
\end{figure*}

The flow structures around the smooth and dimpled airfoils are analyzed using the $Q$-criterion to identify coherent vortical features and gain insight into the unsteady flow dynamics. Figure \ref{fig6} shows the instantaneous iso-surface of $Q$ for $Re_c=5300$ and $Re_c=10,000$, where $Q=0.5 (||\boldsymbol{\Omega}||-||\mathbf{S}||)$ is the second invariant of the velocity gradient tensor which measures the excess of local rotation rate compared to the strain rate \cite{Hunt1988,Jeong_Hussain_1995}. The $Q$-iso-surfaces in the figure are colored by the instantaneous streamwise velocity $U_x^*$ normalized by the freestream velocity to show how fast the corresponding vortical structures move in the $x$-direction. The most dominated vortical structures for both $Re_c$ are located at the leading edge and in the wake regime. For $Re_c=5300$, vortical structures in the wake regime are dominated by the vortex core of the von K\'arm\'an vortex street, shown by the cylindrical vortex structure for all cases in Fig. \ref{fig6}(a)-(c). These vortical structures are similar to those reported in the literature at $Re=5000$ and $\alpha=6.5^\circ$ \cite{Gupta2023}. Three-dimensional effects start appearing around this Reynolds number and  angle of attack, which is consistent with the predictions from stability analysis performed by Gupta et al.\cite{Gupta2023}. The vortical structures are similar for all three cases suggesting a negligible effect of dimples at this Reynolds number, which is consistent with the trends in the force coefficients. At $Re_c = 10,000$, the wake becomes more three-dimensional, as evident from the increased complexity and fragmentation of the vortical structures in Fig.~\ref{fig6}(d)-(f). The emergence of small-scale three-dimensional flow structures is likely related to secondary instabilities, which progressively disrupt the spanwise coherence of the vortex street and lead to enhanced wake turbulence \cite{Gupta2023}.  The vortex sheet in smooth airfoil shows spanwise variation which is diminished with the dimples.


\begin{figure*}[htbp]
  \centerline{\includegraphics[scale = 0.42]{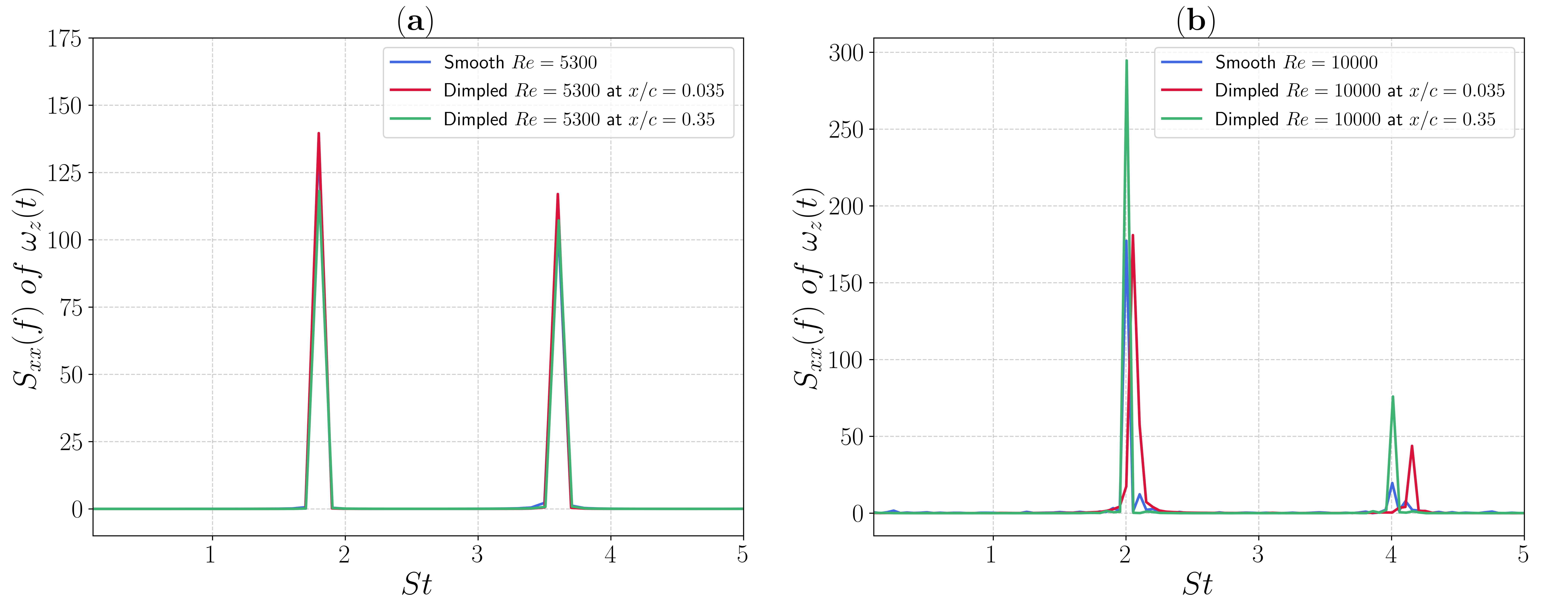}}
  \caption{Power Spectral Density $S_{xx}(f)$ of the instantaneous spanwise vorticity $\omega_z (t)$ at (a) $Re_c=5300$ and (b) $Re_c=10,000$ }
\label{fig7}
\end{figure*}

The two-dimensional side view of the instantaneous normalized spanwise vorticity $\omega_z^*$ is analyzed to examine the vortex street structure in the wake, where $\omega_z^* = \omega_z c / U_\infty$, with $\omega_z = \partial u_y / \partial x - \partial u_x / \partial y$. Spanwise vorticity contours are shown below each $Q$-iso-surface in Fig.~\ref{fig6}. For $Re_c = 5300$, the wake exhibits a well-defined von Kármán vortex street with strong circular vortex cores ($\omega_z^* = \pm 10$), and no significant differences are observed between the smooth and dimpled airfoils (Fig. \ref{fig6}(a)-(c)). This aligns with the minimal effect of dimples on force coefficients. At $Re_c = 10,000$, the wake of the smooth airfoil shows signs of vortex breakdown, with irregular vortex cores indicating the onset of three-dimensional instabilities (Fig. \ref{fig6}(d)). In contrast, both dimpled airfoils ($l_D/c = 0.035$ and $l_D/c = 0.35$) maintain coherent vortex cores ($\omega_z^* = \pm 10$), indicating the delay of wake breakdown. This suggests that dimples exert a stabilizing influence on the wake at this Reynolds number. Similar wake-stabilizing effects of dimples have previously been reported for flapping foils operating at low Reynolds numbers~\cite{silwal2025}.

Figure \ref{fig7} compares the power spectral density (PSD) $S_{xx}(f)$ of the instantaneous spanwise vorticity $\omega_z(t)$ for the smooth and dimpled airfoils at $Re_c = 5300$ and $Re_c = 10,000$, where $S_{xx}(f)=|\hat{x}(f)|^2/\tau$ with $\hat{x}(f)$ is the Fourier transform of a point data signal $x(t)$ and $\tau$ is the signal duration. The point probe is located at $(x,y,z)=(1.65, 1.35,0)$. At $Re_c = 5300$, all cases exhibit a dominant peak at the Strouhal number $St = 1.8$, indicating the same primary vortex shedding frequency (Fig.~\ref{fig7}(a)). A secondary peak at $St = 3.6$ corresponds to the second harmonic of the primary mode. While the peak frequencies remain unchanged by adding dimple array, minor variations in power density are observed. At $Re_c = 10,000$, the shedding frequency shifts slightly to $St \approx 2$, with an overall increase in spectral power (Fig.~\ref{fig7}(b)). The dimpled airfoil with $l_D/c = 0.35$ exhibits a pronounced peak at the primary shedding frequency, whereas the $l_D/c = 0.035$ case shows a slightly shifted frequency at $St \approx 2.05$. A distinguishing feature of the smooth airfoil case is the presence of a double-peak structure, with a secondary peak at St = 2.1, indicating the emergence of secondary instabilities in the wake. This observation is consistent with the findings of Gupta et al.~\cite{Gupta2023}, where Floquet analysis revealed similar behavior, albeit for Reynolds numbers restricted to $Re_c < 5300$. By extrapolating the trend defining the minimum angle of attack required for the onset of three-dimensional wake instabilities (Fig.~7 in \cite{Gupta2023}) to higher Reynolds numbers, it can be inferred that the wake transitions to a three-dimensional state at lower angles of attack as the Reynolds number increases. This secondary peak is absent in the dimpled cases, where only the primary vortex shedding mode remains dominant. A higher harmonic at $St \approx 4$ (twice the primary frequency) is observed across cases, consistent with vortex shedding dynamics. These results align with the observations in Fig.~\ref{fig6}, indicating that dimples influence vortex shedding by strengthening the vortex cores and delaying vortex breakdown in the wake.

\begin{figure*}[htbp]
  \centerline{\includegraphics[scale = 0.26]{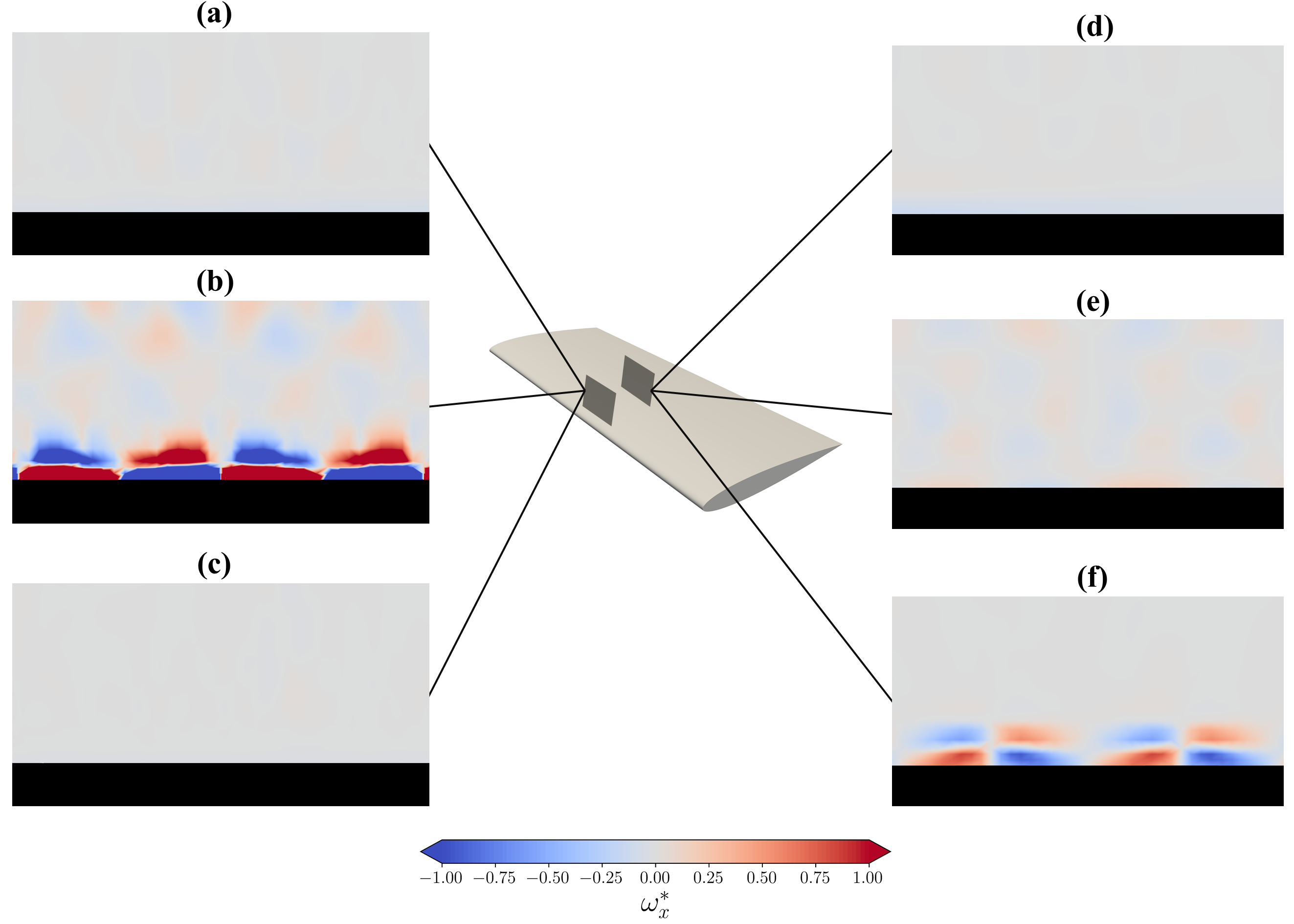}}
  \caption{Close-up view of the instantaneous normalized streamwise vorticity $\omega_x^*$ of $Re_c=10000$ at (a)$-$(c) $x/c=0.065$ and (d)$-$(f) $x/c=0.38$ where (a) smooth airfoil at $t^*=90$, (b) dimpled airfoil at $l_D/c=0.035$ at $t^*=50$, (c) dimpled airfoil at $l_D/c=0.35$ at $t^*=90$, (d) smooth airfoil at $t^*=90$, (e) dimpled airfoil $l_D/c=0.035$ at $t^*=50$, and (f) dimpled airfoil $l_D/c=0.35$ at $t^*=90$. The black color denotes the airfoil body.}
\label{fig9}
\end{figure*}

To further examine the local flow dynamics introduced by surface dimples, Fig.~\ref{fig9} presents the instantaneous normalized streamwise vorticity, defined as \( \omega_x^* = \omega_x c / U_\infty \), where \( \omega_x = \partial u_z / \partial y - \partial u_y / \partial z \), at two downstream locations for \( Re_c = 10{,}000 \): \( x/c = 0.065 \) (Fig.~\ref{fig9}(a)--(c)) and \( x/c = 0.38 \) (Fig.~\ref{fig9}(d)--(f)). At \( x/c = 0.065 \), distinct streamwise counter-rotating vortices are observed for the dimpled airfoil with \( l_D/c = 0.035 \), close to the airfoil surface immediately downstream of the dimple array reaching peak values of \( \omega_x^* = \pm 10 \). As expected these vortices are not seen in the smooth and \( l_D/c = 0.35 \) configurations at this $x/c$ location. Further downstream at \( x/c = 0.38 \), a similar pair of counter-rotating vortices is evident only for the \( l_D/c = 0.35 \) case, while the other two configurations do not exhibit such structures. This suggests that the formation and persistence of these streamwise vortices are strongly dependent on dimple placement and their influence on boundary layer dynamics. Similar streamwise vortices have been reported in earlier studies involving dimpled surfaces at higher Reynolds numbers~\cite{Ligrani2001, Won2005,Xie2022}. Additionally, a secondary, weaker pair of counter-rotating vortices appears above the stronger near-wall structures, oriented with opposite vorticity. The total thickness of this double vortex system is approximately \( 0.023c \), which corresponds to roughly twice the local boundary layer thickness at \( x/c = 0.06 \), and about half that at \( x/c = 0.38 \). Based on these observations, it is hypothesized that streamwise vortices induced by the dimples enhance momentum mixing in the spanwise direction, which may stabilize the wake by suppressing large-scale three-dimensional instabilities. 

\begin{figure*}[htbp]
  \centerline{\includegraphics[scale = 0.26]{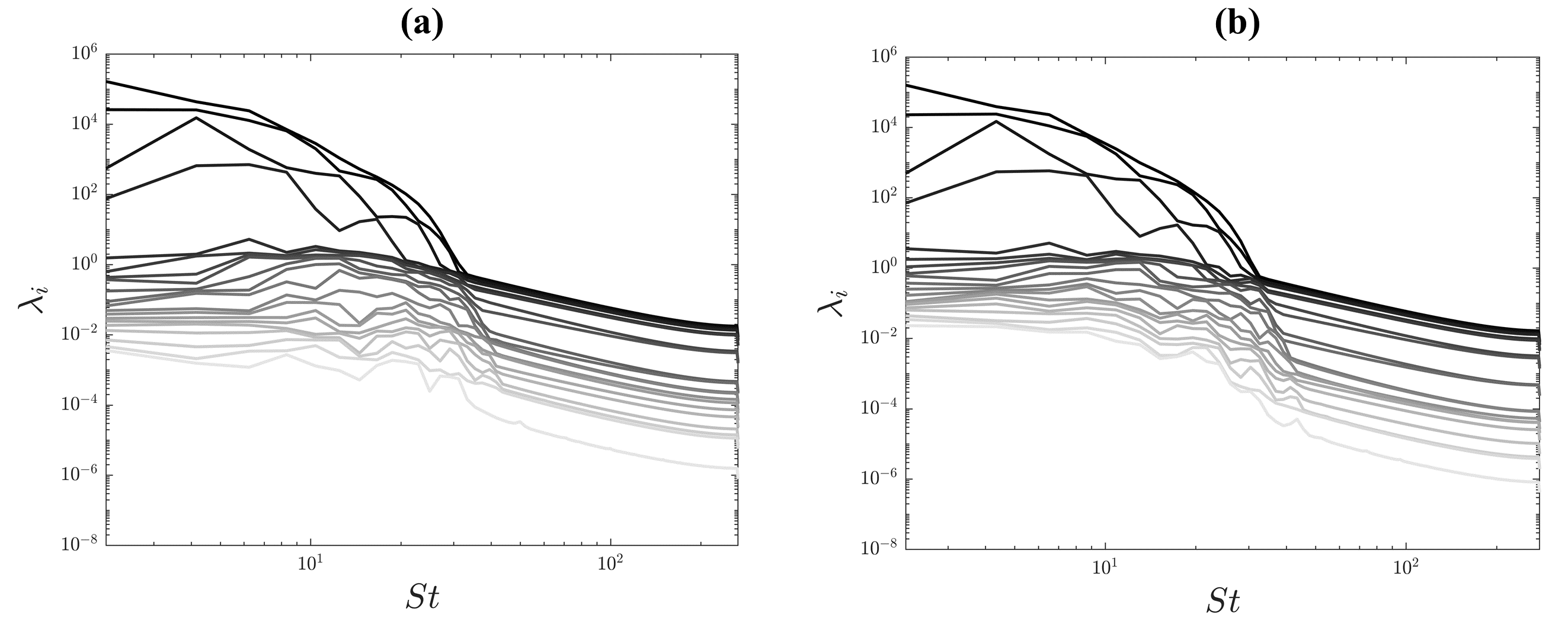}}
  \caption{SPOD energy spectrum $\lambda_i$ for $Re=5300$ of (a) smooth airfoil and (b) dimpled airfoil at $l_D/c=0.35$}
\label{fig10}
\end{figure*}

\begin{figure*}[htbp]
  \centerline{\includegraphics[scale = 0.26]{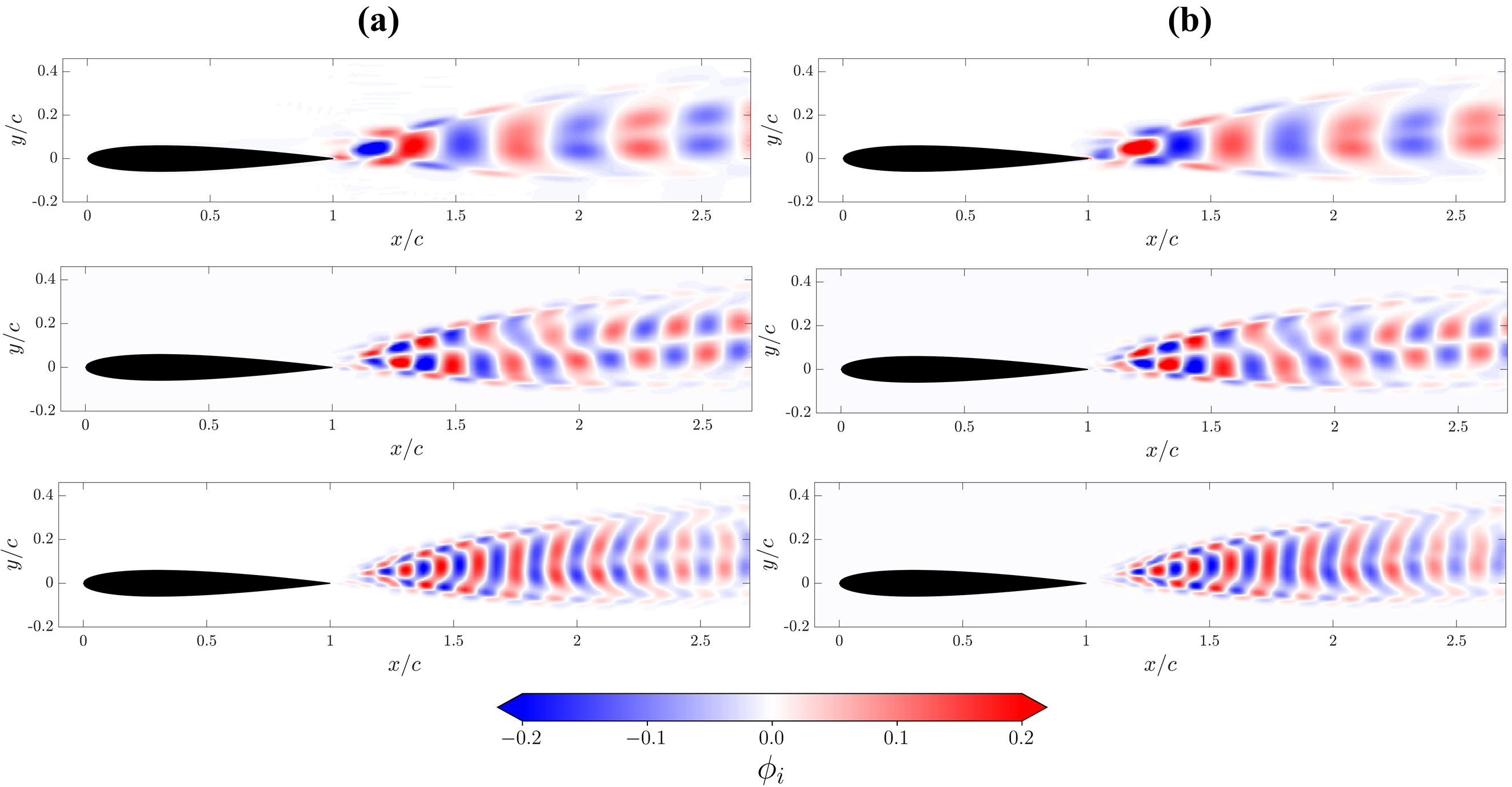}}
  \caption{SPOD modes $\phi_i$ for $Re=5300$ at $St\approx2$ of (a) smooth airfoil and (b) dimpled airfoil at $l_D/c=0.35$. First, second, and third rows correspond to the first $\phi_1$, second $\phi_2$, and third $\phi_3$ modes, respectively.}
\label{fig11}
\end{figure*}

\begin{figure*}[htbp]
  \centerline{\includegraphics[scale = 0.26]{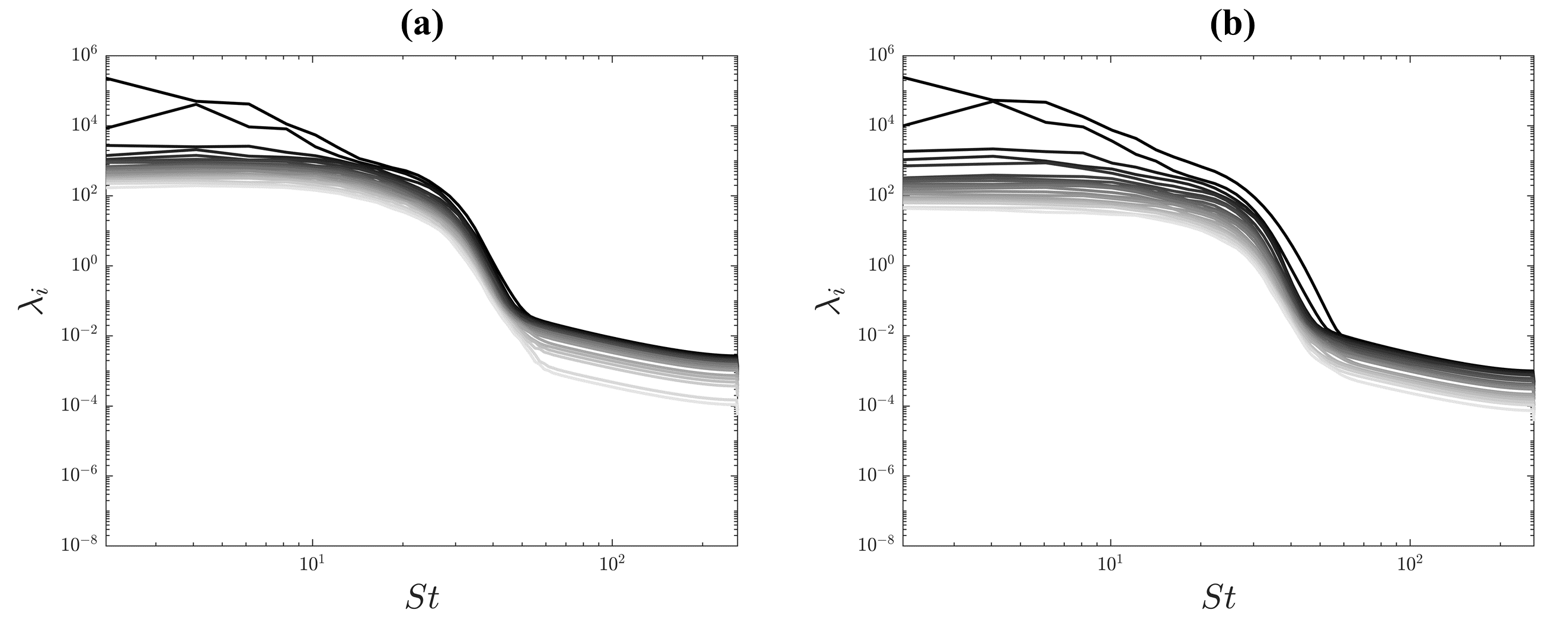}}
  \caption{SPOD energy spectrum $\lambda_i$ for $Re=10,000$ of (a) smooth airfoil and (b) dimpled airfoil at $l_D/c=0.35$}
\label{fig12}
\end{figure*}

\begin{figure*}[htbp]
  \centerline{\includegraphics[scale = 0.65]{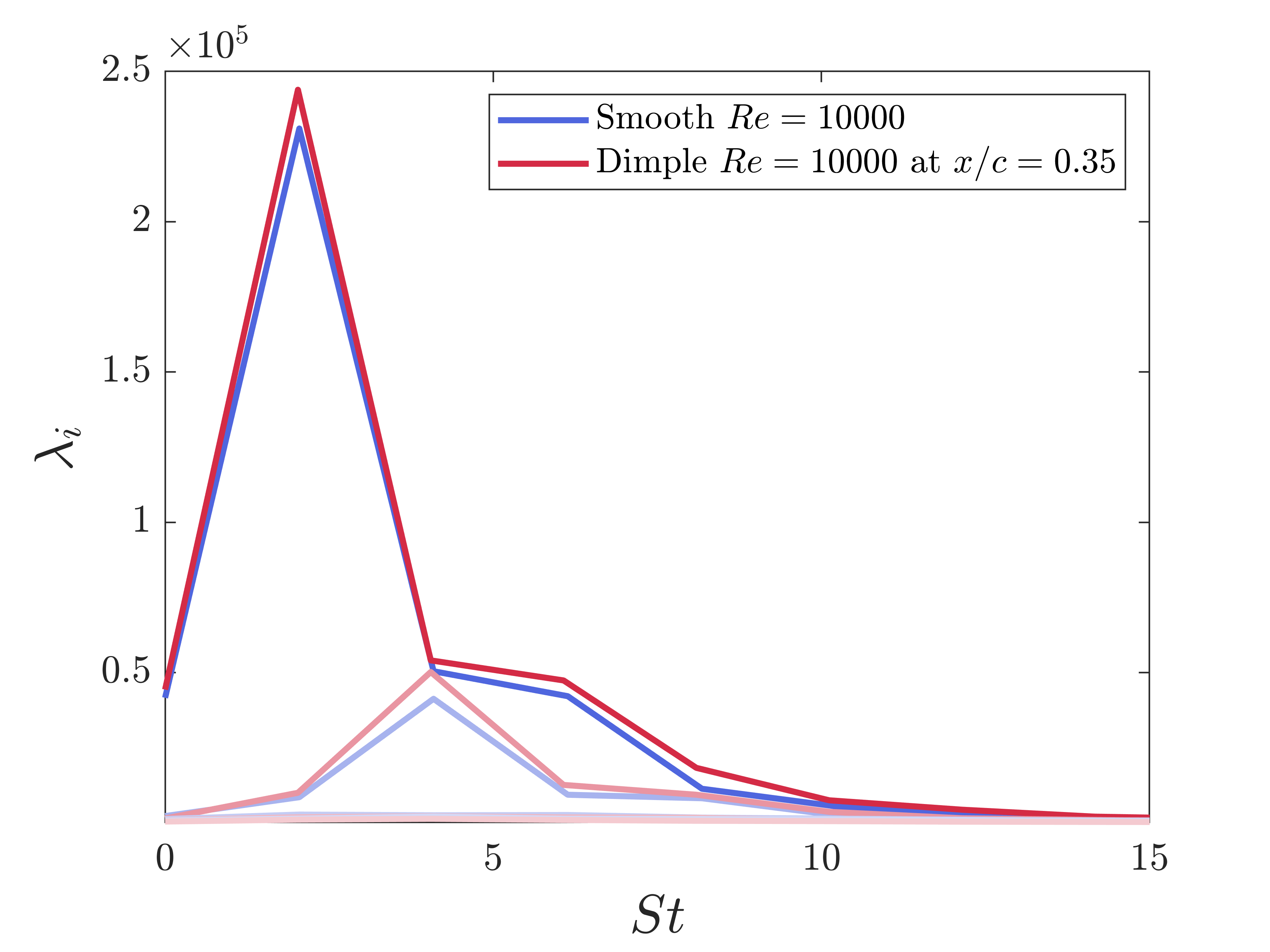}}
  \caption{Largest SPOD energy spectrum $\lambda_i$ for $Re=10,000$}
\label{fig13}
\end{figure*}

\begin{figure*}[htbp]
  \centerline{\includegraphics[scale = 0.26]{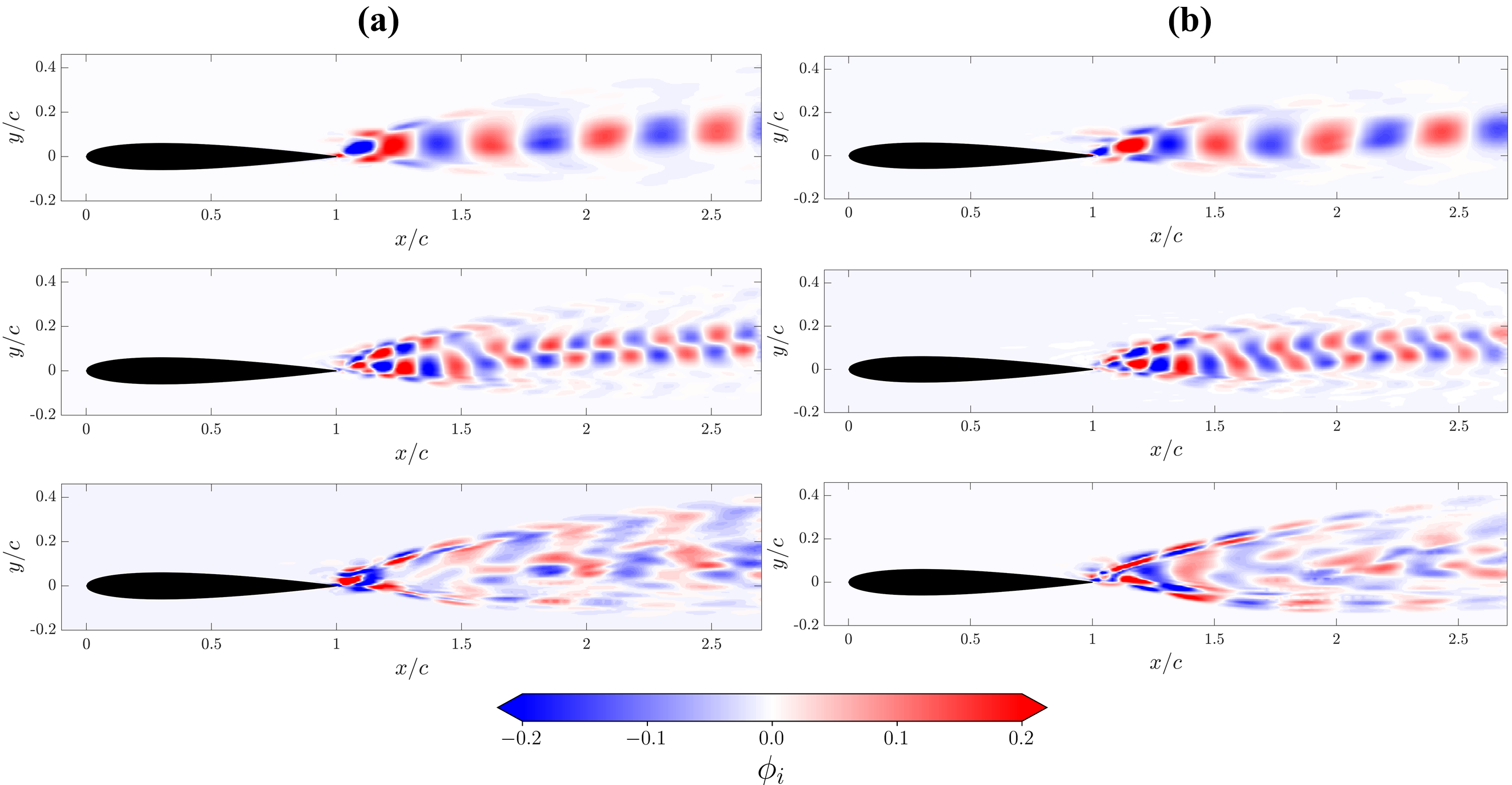}}
  \caption{SPOD modes $\phi_i$ for $Re=10,000$ at $St\approx2$ of (a) smooth airfoil and (b) dimpled airfoil at $l_D/c=0.35$. First, second, and third rows correspond to the first $\phi_1$, second $\phi_2$, and third $\phi_3$ modes, respectively.}
\label{fig14}
\end{figure*}
To further investigate the presence of secondary instabilities in the wake, Spectral Proper Orthogonal Decomposition (SPOD) is performed on the instantaneous spanwise vorticity $\omega_z^*$ in the midspan plane of the foil. SPOD identifies coherent flow structures at a given frequency by performing an eigen-decomposition of the cross-spectral density (CSD) tensor. The resulting SPOD modes correspond to the eigenvectors $\phi_i$, and their associated modal energies are given by the eigenvalues $\lambda_i$. This study follows the algorithm and methodology outlined in prior work \cite{Schmidt2019, Schmidt2020}, with detailed theoretical background available in \cite{Towne2018}. Figure~\ref{fig10} shows the SPOD energy spectra at $Re_c = 5300$ for smooth airfoil and the dimpled configuration with $l_D/c = 0.35$. In both cases, the leading SPOD modes exhibit a pronounced spectral peak at the same Strouhal number, consistent with globally periodic shedding behavior. However, subtle differences emerge in the higher-order modes. While the first two modes are nearly identical between the two configurations, the smooth airfoil displays a broader distribution of energy among the higher modes. In contrast, the dimpled configuration shows a narrower band of elevated energy at these higher modes, suggesting reduced modal complexity and a more organized wake structure.

Figure~\ref{fig11} presents the spatial structure of the three most energetic SPOD modes at \(St \approx 2\) for the smooth airfoil (left column) and the dimpled foil with \(l_D/c = 0.35\) (right column), both at \(Re = 5300\). The first row shows the leading mode \(\phi_1\), followed by the second (\(\phi_2\)) and third (\(\phi_3\)) modes in the second and third rows, respectively. The dominant SPOD mode \(\phi_1\) exhibits a globally coherent shedding structure in both cases; however, a distinct phase shift is observed between the two configurations, with the smooth and dimpled modes appearing approximately \(180^\circ\) out of phase. The phase shift may reflect differences in the timing of vortex formation between the two surfaces. The second and third modes (\(\phi_2\) and \(\phi_3\)) display qualitatively similar spatial features across both cases, indicating that the primary instability mechanisms that govern wake instability are largely preserved. 

Figure~\ref{fig12} presents the full SPOD energy spectra \(\lambda_i\) for the smooth (a) and dimpled (b) configurations at \(Re_c = 10{,}000\). Compared to the lower Reynolds number case (\(Re_c = 5300\)), the spectra exhibit significantly higher energy in the higher-order modes, indicating the presence of more complex, multiscale interactions in the wake at elevated Reynolds number. In both configurations, the first two modes dominate the low-frequency range (\(St \lesssim 4\)) and decay rapidly with increasing frequency. However, the dimpled foil exhibits consistently higher energy in these leading modes.

This trend is further illustrated in Figure~\ref{fig13}, which compares the first and second SPOD eigenvalues, \(\lambda_1\) and \(\lambda_2\), as functions of Strouhal number \(St\) on a linear scale. In addition to the higher energy in the dominant modes, the dimpled case shows a broader and more distributed energy content among the higher modes. This behavior contrasts with the observations at \(Re_c = 5300\), where the dimples led to a narrowing of energy content across modes and a more confined modal structure, coinciding with reduced force fluctuations. It is therefore reasonable to infer that the differing energy distributions across Reynolds numbers contribute to the observed differences in unsteady force behavior. At \(Re_c = 5300\), dimples appear to suppress broadband unsteadiness by concentrating energy into a narrow range of coherent modes, leading to decreased force fluctuations. In contrast, at \(Re_c = 10{,}000\), dimples both redistribute energy into higher-order modes and amplify the energy of the dominant modes. This may explain the observed seven-fold increase in the RMS of force fluctuations and the emergence of highly periodic unsteady loading at the higher Reynolds number.

Figure~\ref{fig14} shows the spatial structures of the three most energetic SPOD modes at \(St \approx 2\) for the smooth airfoil (a) and the dimpled configuration (b) at \(Re_c = 10{,}000\). In both configurations, the leading mode \(\phi_1\) captures the dominant wake instability, characterized by alternating red-blue lobes consistent with vortex shedding. The spatial organization of \(\phi_1\) is broadly similar across both cases, but with a key distinction: the dimpled foil exhibits a phase shift of approximately \(180^\circ\) relative to the smooth foil. A similar out-of-phase relationship is observed for the second mode \(\phi_2\), while the third mode \(\phi_3\) shows slight variations in spatial structure, particularly near the trailing edge and in the early wake region.

These results suggest that while the dominant flow mechanisms primarily global shedding dynamics are preserved, the dimples may induce a systematic shift in the temporal evolution of the coherent structures. This phase offset may arise from a modification of the vortex roll-up timing, or convective velocity due to surface roughness. The presence of dimples appears to introduce a temporal reorganization of the wake without disrupting the underlying modal structure. Unlike the lower Reynolds number case where dimples act to stabilize the flow and reduce unsteady loading, at higher \(Re\) they appear to enhance the strength and coherence of dominant shedding modes, resulting in a highly periodic force response and an approximate seven-fold increase in RMS force fluctuations. 

\section{Summary and Conclusions}\label{sec4}

This study examined the effect of surface dimples on the unsteady aerodynamics of a NACA0012 airfoil at a constant angle of attack \(\alpha = 5^\circ\), using direct numerical simulations at Reynolds numbers \(Re_c = 5300\) and \(Re_c = 10{,}000\). Arrays of dimples were placed at two streamwise locations, \(l_D/c = 0.035\) and \(l_D/c = 0.35\), and their effects were compared to those of a smooth airfoil. Results show that while the mean lift and drag coefficients remained largely unchanged, significant differences were observed in the unsteady force response. At \(Re_c = 5300\), the dimpled configuration with \(l_D/c = 0.35\) yielded a 26.5\% reduction in lift fluctuations and a 33.3\% reduction in drag fluctuations compared to the smooth foil. In contrast, at \(Re_c = 10{,}000\), the same dimple configuration led to a dramatic increase in unsteady forces to a seven-fold increase in the RMS of force fluctuations despite having minimal effect on the mean forces.  Force coefficient time traces at \(Re_c = 10{,}000\) revealed irregular, aperiodic behavior for the smooth airfoil, indicative of a transition to turbulence and the emergence of broadband wake dynamics. In contrast, the dimpled foil at the same Reynolds number exhibited highly periodic force signals, similar to those observed at the lower Reynolds number. This suggests that dimples not only stabilize the dominant unsteady modes but also suppress wake irregularities, delaying the onset of vortex breakdown.

Spectral analysis via power spectral density (PSD) confirmed the presence of a secondary mode near the dominant peak at \(St \approx 2\) for the smooth airfoil at \(Re_c = 10{,}000\), suggesting a secondary wake instability. For the dimpled foil, this secondary wake instability was absent. SPOD analysis further revealed that, at \(Re_c = 5300\), dimples redistributed energy into a narrower band of higher modes, likely contributing to the observed reduction in force fluctuations. However, at \(Re_c = 10{,}000\), dimples increased the energy in the dominant wake modes while also leading to a broader energy spectrum across higher-order modes  presumably causing both the increase in force fluctuations and the enhanced periodicity.

Flow visualizations indicated that counter-rotating streamwise vortices, of the order of the boundary layer thickness, were generated at the trailing edge of the dimples from \(x/c = 0.035\) to \(x/c = 0.2\). These vortices introduced localized perturbations that likely modified the boundary layer dynamics and contributed to the observed changes in unsteady loading. \(Q\)-criterion analysis and instantaneous spanwise vorticity fields at \(Re_c = 10{,}000\) showed increased three-dimensionality and evidence of vortex core breakdown in the smooth foil. In contrast, the dimpled configuration maintained stronger, more coherent vortex cores, suggesting a stabilizing effect of dimples that delayed vortex breakdown.

While dimples are traditionally associated with drag control on bluff bodies, this study demonstrates their potential effectiveness on streamlined geometries as well. The ability of dimples to modulate unsteady forces, suppress vortex breakdown, and enhance coherent structures highlights their promise for use in flow control of airfoil-based energy harvesting systems or unmanned aerial vehicles. Future work will extend this investigation to higher Reynolds numbers to explore the broader applicability of dimples in turbulent regimes.

\section{Acknowledgments}\label{sec5}
We acknowledge Great Lakes High-Performance Computing (HPC) at the University of Michigan, Ann Arbor. We are also grateful to the Indonesia Endowment Fund for Education (LPDP) under the Ministry of Finance, Republic of Indonesia for their financial support.

\bibliographystyle{elsarticle-num} 
\bibliography{refdimple}

\end{document}